\documentclass[10pt,aps,prl,twocolumn,superscriptaddress,showpacs]{revtex4-1}
\usepackage{amsmath}
\usepackage{amsfonts}
\usepackage{graphicx} 
\usepackage{float}
\usepackage{dcolumn}
\usepackage{bm}
\usepackage{color}

\newcommand*\xbar[1]{%
   \hbox{%
     \vbox{%
       \hrule height 0.5pt 
       \kern0.3ex
       \hbox{%
         \kern-0.1em
         \ensuremath{#1}%
         \kern-0.1em
       }%
     }%
   }%
}

\newcommand{\varphivec}{\mbox{\boldmath$\varphi$}}
\newcommand{\etavec}{\mbox{\boldmath$\eta$}}
\newcommand{\br}{{\mathbf{r}}}

\newcommand{\be}{\begin{equation}}
\newcommand{\ee}{\end{equation}}
\newcommand{\avgw}{{\langle w\rangle}}
\newcommand{\bea}{\begin{eqnarray}}
\newcommand{\eea}{\end{eqnarray}}

\newcommand{\argu}[1]{\left({#1}\right)}
\newcommand{\argz}[1]{\left[{#1}\right]}
\newcommand{\args}[1]{\left\{{#1}\right\}}

\begin{document}

\title{Classical Theory of Quantum Work Distribution in Chaotic Fermion Systems}

\author{Andr\'as Grabarits}
\affiliation{BME-MTA Exotic Quantum Phases `Lend\"ulet' Research Group, Institute of Physics, Budapest University of Technology and Economics, 
Budafoki \'ut 8., H-1111 Budapest, Hungary}
\affiliation{MTA-BME Quantum Dynamics and Correlations Research Group, 
Institute of Physics, Budapest University of Technology and Economics, 
Budafoki \'ut 8., H-1111 Budapest, Hungary}
\author{M\' arton Kormos}
\affiliation{MTA-BME Quantum Dynamics and Correlations Research Group, 
Institute of Physics, Budapest University of Technology and Economics, 
Budafoki \'ut 8., H-1111 Budapest, Hungary}
\author{Izabella Lovas}
\affiliation{Department of Physics and Institute for Advanced Study,
Technical University of Munich, 85748 Garching, Germany}
\affiliation{Munich Center for Quantum Science and Technology (MCQST), Schellingstr. 4, D-80799 M\" unchen}
\author{Gergely Zar\'and}
\affiliation{BME-MTA Exotic Quantum Phases `Lend\"ulet' Research Group, Institute of Physics, Budapest University of Technology and Economics, 
Budafoki \'ut 8., H-1111 Budapest, Hungary}
\affiliation{MTA-BME Quantum Dynamics and Correlations Research Group, 
Institute of Physics, Budapest University of Technology and Economics, 
Budafoki \'ut 8., H-1111 Budapest, Hungary}

\date{\today}

\begin{abstract}
We present a theory of quantum work statistics in generic chaotic, disordered Fermi liquid systems within a driven random matrix formalism. 
By extending P. W. Anderson's orthogonality determinant formula to compute quantum work distribution, we find that 
 work statistics is non-Gaussian and is characterized by a few dimensionless parameters.  At longer times,
quantum interference effects become irrelevant and the quantum work distribution is well-described in terms of a purely 
classical ladder model with a symmetric exclusion process in energy space, 
while bosonization and mean field  methods provide accurate analytical expressions for the work statistics.
Our random matrix and mean field predictions are validated by numerical simulations for  a  two-dimensional disordered quantum  dot, 
and can be  verified  by calorimetric measurements on nanoscale circuits.
\end{abstract}

\maketitle

\paragraph{Introduction.---} 

The concepts of heat and work lie at the foundations of thermodynamics and statistical physics. When considered in the quantum realm, however, they raise deep questions and pose new challenges \cite{workreview}. Even the very definitions of heat and energy transfer become nontrivial as they require the specification of the measurement protocol \cite{Hanggi}.  At the same time, the interplay of quantum and thermal fluctuations, coherence, and dissipation gives birth to novel phenomena which are in the focus of the rapidly growing field of quantum thermodynamics connecting quantum physics, thermodynamics, and quantum information theory \cite{goold,otocGu}. With the recent experimental developments, these issues are not purely academic but can be studied in the laboratory, in systems ranging from individual molecules \cite{bioreview,molecule1,molecule2}  through mesoscopic grains~\cite{pekola,pekola2} and nuclear spins~\cite{batalhao} to cold atoms~\cite{cerisola} and nitrogen vacancy centers~\cite{NVC}.

The definition and measurement of work  in quantum systems requires a two-time measurement protocol:  one first
determines   the energy $E^i_{0}$ of the initial state at time $t=0$, and later, in a second measurement, the 
energy $E^f_{t}$  of the time evolved system at time $t$. The adiabatic part being essentially trivial, here we focus on  the 
`entropic' contribution of energy absorption or `work',  defined as $W \equiv E^f_{t} - E^i_{t}$, i.e., the energy absorbed ($W>0$) 
or emitted ($W<0$) by  the system due to non-adiabatic  transitions, and investigate the corresponding distribution function, $P_{\,t}(W)$. 
The full distribution of work has been studied extensively in many-body systems ~\cite{silva,gambassi,lutt,FD,FD1,FD2,liebliniger},
and its characteristic function of this distribution has been related to the Loschmidt echo \cite{silva,chenu} and to quantum information scrambling \cite{chenu}. However, the effect of disorder and randomness is much less studied \cite{random1,random2,random3} despite their relevance in mesoscopic systems.

\begin{figure}[b!]
\includegraphics[width=0.9\columnwidth]{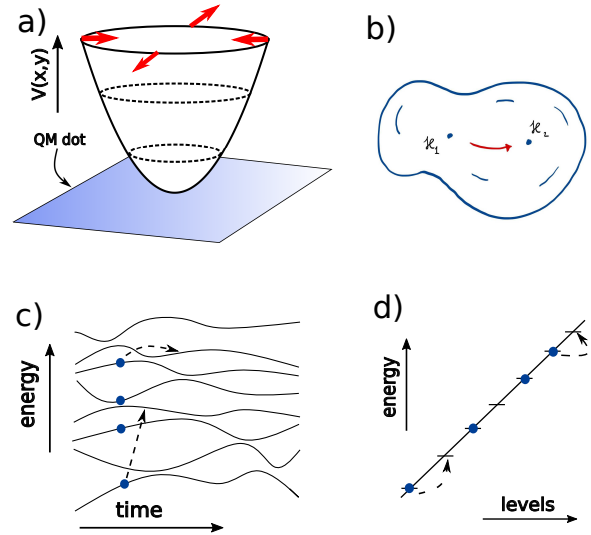}
\caption{
a) Disordered  2-dimensional electron gas  with a parabolic potential deformed in time, driving the 
system away from equilibrium. b) Motion in the manifold of random matrices.
c)  Deformation-induced motion  of  energy levels,  giving rise to  particle-hole excitations.
d) `Ladder' model: classical diffusion of hard core particles between  uniformly spaced energy levels.}
\label{fig:sketch}
\end{figure}

To fill this gap, here we  focus on  disordered, chaotic fermion systems such as 2-dimensional quantum dots, 
which we perturb by changing  external gate voltages, fields, and electrodes, as shown in 
Fig.~\ref{fig:sketch}.a.   We neglect
interactions under the assumptions that a  non-interacting Fermi liquid description is appropriate. 
Under these conditions, the system can be described  in terms of  the time dependent Hamiltonian
 \begin{equation}
\hat{H}(t)=\sum_{i,j=1}^N \hat{a}_i^\dagger\,\mathcal{H}_{ij}(t)\,\hat{a}_j\,,
 \label{eq:H}
\end{equation}
where  the $\hat{a}_i$'s stand for  fermionic annihilation operators, and the single particle Hamiltonian  $\mathcal{H}(t)$ 
incorporates  disorder effects and also accounts for the impact of time dependent electrodes.
The total fermion number is conserved by Eq.~\eqref{eq:H}, $\sum_i  \hat{a}_i^\dagger\hat{a}_i = M$.
 For a concrete physical system such as a
quantum dot defined in  a disordered 2 dimensional electron gas, we can and will construct  microscopic models for 
$\mathcal{H}(t)$  and compute  work statistics.  
The single particle spectrum of most chaotic systems is, however,  known to be  captured by random matrix theory~\cite{matrixreview,RMreview}. 
 We  can therefore also follow the strategy of  Refs.~\cite{wilkinson2}  and \cite{PRR}, and  consider deformations within 
the space of  Gaussian random matrix ensembles, 
$$
\mathcal{H}(t)=\mathcal{H}_1\cos\lambda(t)+\mathcal{H}_2\sin\lambda(t)\;,
$$
with  $\mathcal{H}_{1,2}$  some independent $N\times N$ Gaussian  matrices from the orthogonal (GOE), unitary (GUE) 
or symplectic (GSE) ensembles, and $\dot{\lambda} = v $ setting the speed of deformations.  In this latter case, 
the parameter $\lambda$ generates a motion along an 'arc' or 'circle' within the random matrix ensemble, as depicted in
Fig.~\ref{fig:sketch}.b.

Our goal is to understand universal aspects of the structure and time evolution of 
the distribution  $P_t(W)$. For simplicity, here we  focus on \emph{quantum quench} 
protocols, i.e.,  we start  from the ground state of ${\hat H}(0)$, but our results can be readily generalized to finite 
temperature mixed states~\cite{thermal}. We follow the quantum evolution of the 
disordered many-body systems, and use a determinant formula presented in Ref.~\cite{PRR}
to compute $P_t(W)$. We find that  the statistics of $P_t(W)$ is almost  independent 
of microscopic details as well as the symmetry of the Hamiltonian,
once the absorbed energy exceeds sufficiently the one-body energy separation $\delta\epsilon \equiv  1/N(\epsilon_F)$, characterizing the total density of levels 
at the Fermi energy $\epsilon_F$, and the time is long enough, $t>\hbar/\delta\epsilon$. To capture work  in this long time limit, we construct 
a classical `ladder' model which incorporates quantum statistics as 
well as level repulsion, but ignores interference effects between consecutive level collisions and Landau-Zener transitions.
Our `ladder' model  gives a surprisingly accurate description of $P_t(W)$, and allows us to derive 
accurate analytical approximations for $P_t(W)$ by means of  bosonization and  a particle number conserving 
mean field method. We also validate the RMT description and the `ladder' model in a 2D quantum dot system.

\paragraph{Quantummechanical analysis.---} 
Since the Hamiltonian $H$ is non-interacting, all information is contained in the time evolution of 
the single particle wave functions, $\varphivec^m(t)$. These can be obtained by expanding 
$\varphivec^m(t)$ in terms of the instantaneous eigenfunctions $\etavec^k_t$ of $\cal H$,   
as $\varphivec^m(t) =\sum_k\alpha^m_k(t) \,\etavec^k_t$, and then solving the single particle 
Schr\"odinger equation for $\alpha^m_k(t)$.   The 
generating function $G_t(u)$ of the work distribution $P_t(W)$ can then be expressed by a simple determinant formula ($\hbar =1$)~\cite{PRR,FeiQuan}
 \begin{align}
G_t(u)&=
\big\langle  \big\langle\Psi(t)|\, e^{iu\left(\hat{H}(t)-E_\text{GS}(t)\right)}\,|\Psi(t)\big\rangle\big\rangle_\text{RM}\nonumber\\
&=\big\langle e^{-i\,u\sum_{m=1}^{M}\varepsilon_m(t)}\,{\rm det}\, g_t(u)\big\rangle_\text{RM}\;,
\label{eq:Gu}
\end{align}
where the matrix $g_t(u)$ contains information on  overlaps  and the instantaneous single particle energies $\varepsilon_k(t)$  at time $t$, 
$\left[g_t(u)\right]^{m m^\prime}\equiv \sum_k [\alpha_k^m(t)]^*\,e^{i\,u\,\varepsilon_k(t)}\,\alpha_k^{m^\prime}(t)$.
We compute $g_t(u)$ numerically,  average over disorder or the random matrix ensemble, $\langle\dots\rangle_{\mathrm{RM}}$, and determine the final distribution   by performing a Fourier transformation. 

The spacing $\delta\epsilon$ and its inverse provide natural energy and time scales, and allow  
us to introduce the dimensionless  work and time, $w\equiv W/\delta\epsilon$ and $\tilde t \equiv t\,\delta\epsilon$, respectively. 
As shown in Fig.~\ref{fig:sketch}.c, deformations of the Hamiltonian lead to a continuous motion of 
single particle levels, and thereby induce collisions and transitions between them.  These 
collisions and Landau--Zener transitions give rise to  a  \emph{diffusive} 
broadening of the Fermi surface at longer times, where – after a short time perturbative $\sim t^2 $ 
scaling –  the average work is found to increase as $\langle w\rangle = \widetilde D \,\tilde t$ with 
$\widetilde D$ the dimensionless energy diffusion constant (see Refs.~\cite{SuppMat} and \cite{PRR}). 

 The distribution $P_{\,\tilde t\,}(w)$ can be disentangled into an adiabatic  and a
 regular part,   
\begin{align}\label{eq:preg}
P_{\,\tilde t\,}(w)=P_\text{ad}(\,\tilde t\,)\, \delta(w)+P_{\rm reg}(w;\,\tilde t\,)\;. 
\end{align}
Random matrix theory implies that – apart from the symmetry of the Hamiltonian  –  the statistics of the evolution of the 
eigenvalues, sketched  in Fig.~\ref{fig:sketch}.c, is completely 
characterized  by  the \emph{velocity} with which levels  deform, i.e., the frequency of avoided level crossings. 
Indeed, the  average  distance of level crossings,  $\langle\Delta \lambda\rangle$ and the time scale $1/\delta\epsilon$ define 
 a natural `velocity' in parameter space, $v_c \equiv \langle\Delta \lambda\rangle \delta\epsilon$, which  we can use to introduce the dimensionless 
velocity,   $\tilde v \equiv \dot \lambda / ( \langle \Delta \lambda\rangle \delta\epsilon)$~\cite{footnote_RMTscaling}. 
The dimensionless velocity characterizes microscopic processes. For $\tilde v \ll 1$ the motion is almost adiabatic, and 
small probability Landau--Zener transitions dominate.   For $\tilde v \gg 1$, on the other hand, transitions between remote levels 
generate energy absorbtion.

From our random matrix considerations it follows that the distribution  $P_{\,\tilde t}(w)$   can only depend  on $\tilde t$, $\tilde v$,  and, in case of finite temperature initial states, on the dimensionless initial 
temperature, $\widetilde T \equiv T/\delta\epsilon$. Similarly, the diffusion constant ${\widetilde D}$ is  a universal function of $\tilde v$, which scales 
as $\widetilde D\sim \tilde v ^2$ for large velocities, while for $\tilde v <1$ nearest 
neighbor transitions dominate and yield  $\widetilde D\sim \tilde v ^{(\beta/2 +1)}$, with $\beta = 1,2$ and $4$ characterizing the 
orthogonal, unitary, and symplectic ensembles, resepectively (see the Supplementary Material~\cite{SuppMat}).

\begin{figure}[t!]
\includegraphics[width=0.9\columnwidth,trim={8.2cm 0 12cm 0},clip]{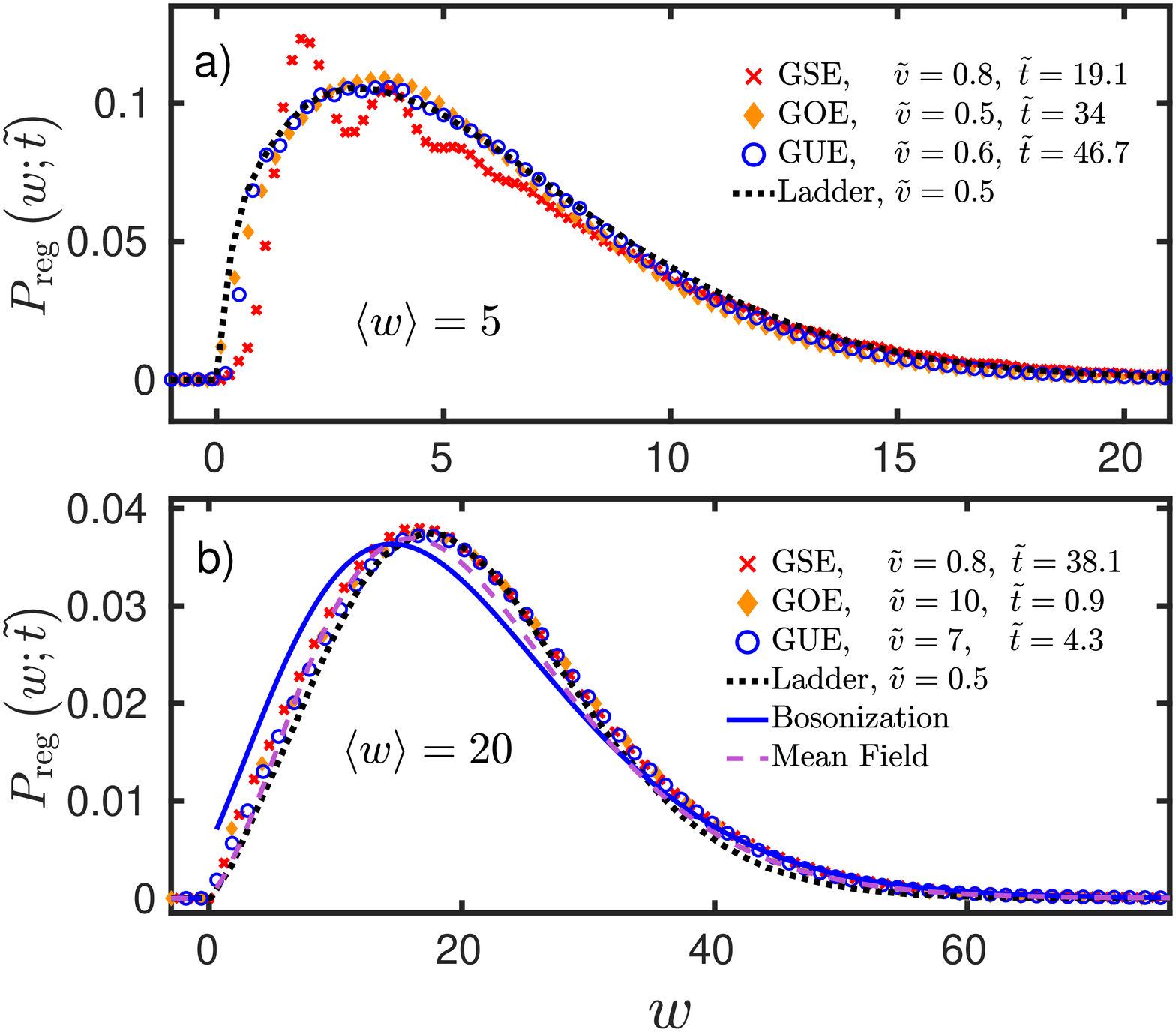}
\caption{Work statistics for GOE, GUE, GSE, for  dimensionless average work $\langle w\rangle=5$ (a),  and  $\langle w\rangle=20$ (b).
For smaller $\langle w\rangle$, $P_{\rm reg}(w;\,\tilde t\,)$ displays features associated with level repulsion and specific to the symmetry of 
the underlying Hamiltonian, while for large  $\langle w\rangle$,
the distributions $P_{\rm reg}(w;\,\tilde t\,)$ fall onto a single curve for all universality classes.  Mean field (dashed line) and bosonization (continuous line) 
approaches give accurate description in the diffusive regime.
}\label{fig:P_reg}
\end{figure}

For small work, $\langle w\rangle \lesssim 10$,  the statistics depends on $\beta$ as well as on $\tilde v$
and $P_{\rm reg}(w;\,\tilde t\,)$ displays peaks and minima  associated with level repulsion,
clearly reflecting the  symmetry of the underlying Hamiltonian    (see Fig.~\ref{fig:P_reg}.a). 
For larger works, $\langle w\rangle \gtrsim \max\, \{\tilde v^2,1\}$, however,  one enters a diffusion dominated regime, where 
symmetry related and microscopic features become less important, and a 
universal distribution displayed  in  Fig.~\ref{fig:P_reg}.b emerges. The observed distribution is clearly non-Gaussian, 
and  characterizes work statistics in generic fermion systems.

\begin{figure}[t!]
\includegraphics[width=1\columnwidth]{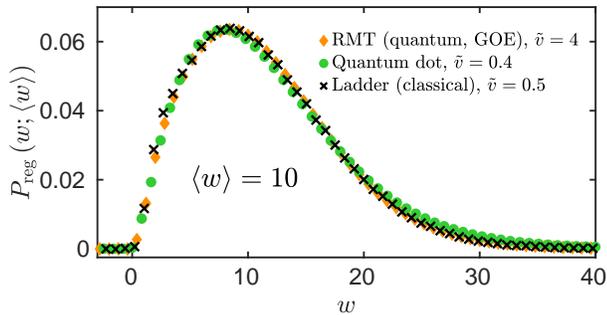}
\caption{
Work statistics for dimensionless average work $\langle w\rangle=10$.   Microscopic   quantum dot (QD) model simulations (green circles),  
 random matrix (GOE) results (orange diamonds), and  the `ladder' model statistics (black crosses) fall on top of each other  with  good accuracy. QD 
calculations were performed for $M=427$ electrons for a lattice of size $38\times38$, disorder $D=1.75J$,
potential strength $\alpha=75J$, and a dimensionless velocity, $\tilde v=0.4$.
For the GOE computations we used $N = 40$ with $M=20$ electrons and a velocity $\tilde v=4$.
For the `ladder' model simulations we used, $\tilde v=0.5$ and  $N=120$ levels with $M=60$ electrons. 
Quantum work distribution depends only on the average work $\langle w\rangle$ and is well captured  
by the classical `ladder' model. 
}
\label{fig:P(W)}
\end{figure}

\paragraph{Ladder model.---} The agreement between the three universality classes  
is suggestive  that quantum interference effects do not play an important role in this diffusion-dominated regime. 
We can therefore attempt and construct a classical \emph{`ladder'} model,   consisting  of uniformly 
placed classical energy levels at a distance $\delta\epsilon$ from each other,  
\be
\epsilon_k = k\,\delta\epsilon \,,\quad\quad 
k=1,2,\dots\,,
\label{eq:ladder}
\ee
occupied by  hard core particles in line with Fermi statistics. The energy of a many-body state is then given by 
$E=\sum_k n_k\,\epsilon_k$ with $n_k \in \{0,1\}$ the occupation numbers, and $\sum_k n_k = M$
the total  number of particles.  The evenly placed levels \eqref{eq:ladder} mimic level repulsion and  level rigidity in 
 chaotic systems. As a final component, perturbation-induced random Landau--Zener transitions are modeled by  nearest neighbor 
hopping transitions and a symmetrical exclusion process (SEP) in energy space.   This simple model captures the diffusive broadening of 
the Fermi surface (see Ref.~\cite{SuppMat}) and, in addition to level repulsion, it also incorporates Fermi statistics and 
particle number conservation.  As can be seen in Figs. \ref{fig:P_reg} and \ref{fig:P(W)}, this classical stochastic model gives a surprisingly accurate description of the work statistics for large enough average work, independently of the velocity.
Moreover, with certain assumptions, the `ladder' model can be used to 
compute $P_\mathrm{\,ad}(\,\tilde t\,)$ and $P_{\rm reg}(w;\,\tilde t\,)$ analytically 
 for a $T=0$ temperature initial state,  without performing  the actual  Monte Carlo simulations, using either bosonization or a more accurate mean field approach. It is, however, crucial to treat particle 
 number conservation with care.

\paragraph{Bosonization.---} 
Bosonization offers a simple method to treat particle number conservation in the `ladder' model. 
Introducing fermion operators for each level, we can express the total energy as $H=\sum_k (\epsilon_k-\epsilon_F)  :c^\dagger_k c_k:$
with $\epsilon_F = \delta\epsilon\,(M+1/2)$ the Fermi energy and $:...: $ referring to normal ordering
with respect to the Fermi sea.   Following Ref.~\cite{vonDelft}, we   introduce 
bosonic operators, $b^\dagger_{q>0}\equiv (1/\sqrt{q})\, \sum_k c^\dagger_{k+q}c_k$, which satisfy  the usual commutation relations, 
$[b_q,b^\dagger_{q^\prime}] = \delta_{q,q^\prime}$, and rewrite the Hamiltonian in terms of these as 
\be
H = \sum_{q\in \mathbb{Z}^+} \delta\epsilon\, q \; b^\dagger_q b_q + \frac {\delta\epsilon} 2 { \hat N}^2
\label{eq:bosonized}
\ee
with $ \hat N  = \sum_k c^\dagger_k c_k -M$ the normal ordered fermion number. 
Clearly, the fermion number does not change for the closed system studied here so the second term in Eq.~\eqref{eq:bosonized}
does not give a contribution. We can obtain an approximate expression for $P_{\,\tilde t}(w)$ by assuming that the final state is thermal with an effective boson 
temperature ${\widetilde T}_\mathrm{eff} = \sqrt{6\langle w\rangle} / \pi$, chosen to yield the appropriate 
average energy,  $\langle  \sum_{q>0}  q \; b^\dagger_q b_q \rangle \equiv \langle w\rangle  $. 
In the large $\langle w\rangle$ limit, we then  obtain (see~\cite{SuppMat}), 
\begin{equation}
P^\mathrm{\,Bose}_{\tilde t}( w) \approx e^{-\frac{\pi^2 {\widetilde{T}}_\mathrm{eff}}  {6}} 
\,  \Bigl[\frac{\pi}{\sqrt{6\, w}}e^{-w/\widetilde T_\text{eff}}
		\;{I}_1\bigl(\pi \sqrt{\textstyle{\frac23}w}\bigr) +\delta\left(w\right)\Bigr],
\label{eq:P_bos}
\end{equation}
where  $I_1$ is the modified Bessel function of the first kind. Since $T_\mathrm{eff}\sim  \sqrt{\widetilde D \tilde t}$, 
the prefactor decays as $\sim e^{- C \sqrt{\widetilde D \tilde t}}$, corresponding to a stretched exponential decay of adiabatic 
processes, as confirmed by our quantum mechanical simulations~\cite{PRR}.

\paragraph{Mean field theory.---} 
The bosonization approach yields a good account of the overall structure of $P_{\tilde t}(w)$, but with certain limitations (see Fig. \ref{fig:P_reg}b). 
In particular, the assumption of a thermal final state is not quite correct. The occupation of the single particle  levels after the time evolution 
is not described by the Fermi function but has a diffusive structure, as stated earlier. 
A more accurate expression can be obtained for $P_{\,\tilde t}(w)$ in a simple, particle number conserving \emph{mean field} approach, 
where instead of assuming   thermalization, we rely on the diffusive nature of energy absorption,
and  assume that each fermion level $k$ is occupied  with probability 
$f_k = (1-\mathrm{erf}[(\tilde \epsilon_k-\tilde\epsilon_F)/\sqrt{4\tilde D \tilde t}\,])/2$, corresponding to a diffusive broadening of the 
Fermi surface. To  enforce the constraint, $\sum_k n_k = M$, we use an integral representation over an auxiliary variable.
A  saddle point procedure in this latter then yields accurate expressions for $P_\text{\,ad}(\,\tilde t\,)$ as well as 
for $P_{\rm reg} (w;\,\tilde t\,)$.

The mean field   probability distribution, $P^{\rm MF}_{\,\tilde t} (w)$, is similar in structure to 
Eq.~\eqref{eq:P_bos}, but  contains additional correction terms  (see Ref.~\cite{SuppMat} for details), 
\begin{multline}
P^\mathrm{\,MF} _{\tilde t}(w) \approx  P_{\,\mathrm{ad}}^\mathrm{\,MF}\; \delta(w)\,  +\\
\frac{c_w}{\sqrt{w}}  \, e^{-c_w \frac{w + \avgw}{\sqrt{\avgw}}}
\left[ 
I_1 (2c_w\sqrt{w})  - \sqrt{2} \,I_1 \bigl( 2c_w\sqrt{ w/2}\bigr)\right]
\end{multline} 
with $c_w\approx 1.35$ and 
\begin{equation}
P^\mathrm{\,MF}_\text{\,ad} = (8\pi\langle w \rangle )^{1/4}\,{e^{-c_w\sqrt{\langle w \rangle}}}\,.
\end{equation}
As shown in Fig.~\ref{fig:P_reg}.b,   the mean field expressions above  yield an  accurate description 
of work in the diffusive regime. Similar to the bosonization result, Eq.~\eqref{eq:P_bos},   $P^{\rm MF}_{\,\tilde t} (w)$
is non-Gaussian and, by construction,  depends parametrically only on $\langle w\rangle$.   
The probability of adiabatic processes also falls off as a stretched exponential, 
but the prefactor $c_w$ is more accurate than the one obtained 
by the simple bosonization theory ($\pi/\sqrt{6}\approx 1.28$)~\cite{SuppMat,PRR}.   

\paragraph{Validation by microscopic models and experimental setup.---}
To confirm the predictions above and to validate the results of our  random matrix approach,
 we propose to study a 2-dimensional quantum dot (QD), and squeeze the electron gas confined 
there by applying time dependent external gate voltages (see Fig.~\ref{fig:sketch}.a).  
This system can be  realized experimentally \cite{calorimetry1,calorimetry2}. 

We model  the QD by a disordered tight binding  Hamiltonian, 
\be
H = - J \sum_{\br,\bm{\delta}} c^\dagger_{\br+\bm{\delta}} \,c_{\br} + \sum_\br (V(\br,t) + \epsilon_\br) c^\dagger_\br c_\br \;,
\label{eq:H_dot}
\ee
where the first term accounts for the kinetic energy of the electrons,  while the potential 
$V(\br,t) = \frac 1 2 ( \alpha r^2 + \lambda(t) (x^2-y^2) )$ describes  the  parabolic confinement, generated by 
external gate electrodes. The second term in $V(\br,t)$ describes a 
compression (decompression) of the electron gas  in the $x$ direction  with a simultaneous 
decompression (compression) along the $y$ direction. We vary $\lambda$ to  induce 
 deformations and generate dissipation. Finally, the random onsite energies  $\epsilon_\br$ are drawn from a 
Gaussian distribution of variance, and are responsible for  electron scattering and disorder.   

A numerical investigation  of the single particle spectrum of Eq.~\eqref{eq:H_dot} reveals that, although some deviations are clearly present, 
the spectrum of   Eq.~\eqref{eq:H_dot}  is reasonably described in terms of GOE for each value of $\lambda$ (see Ref.~\cite{SuppMat}). 
We generate work then by  varying  $\lambda$ uniformly in time,  and use the determinant formula in Eq. \eqref{eq:Gu} 
to compute $P_{\tilde t}(w)$. The disorder-averaged 
results  for $P_{\mathrm{reg}}(w;\,\tilde t\,)$ are  presented for $\langle w\rangle = 10 $ in Fig.~\ref{fig:P(W)}. They show striking 
agreement with random matrix theory as well as with the `ladder' model, and thereby validate the latter.

An altenative experimental platform to study quantum work statistics is offered by ultracold atoms \cite{cerisola}. 
For a forward-backward protocol $P_\mathrm{ad}$ is essentially the ground state fidelity,  which  has been measured in 
Ref.~\cite{kaufman} by preparing  two identical copies of a quantum system, and measuring their overlap. 
This method could be used to verify the predicted stretched exponential behavior of $P_\mathrm{ad}$
in disordered fermion systems. 

\paragraph{Conclusions.---}
We studied the full distribution of quantum work in disordered non-interacting fermion systems both within the framework of random matrix theory and in concrete microscopic model. Surprisingly, we found that for large enough average work, the distribution is independent of the random matrix ensemble and is very well captured by a classical stochastic model describing diffusion in energy space. This allowed us to make various simplifications (e.g. `ladder' model) and derive approximate analytic expressions via bosonization and mean field theory. Interestingly, the bosonization result in Eq. \eqref{eq:P_bos} also emerged in the context of work statistics in Luttinger liquids after an interaction quench~\cite{lutt}. Let us stress that the final state is not thermal but has a diffusive occupation profile, which is the reason why the bosonization approach performs more poorly in comparison with the mean field treatment (cf. Fig. \ref{fig:P_reg}b). For an experimental realization, we propose to study squeezed disordered quantum dots where our results could be tested experimentally.

\paragraph{Acknowledgments.---} We thank Adolfo del Campo for interesting discussions.  
This work was supported by the National Research, Development and Innovation Office (NKFIH)
through  the Hungarian Quantum Technology National Excellence Program, 
project no. 2017-1.2.1-NKP-2017- 00001,  and by 
the \'UNKP-20-5 New National Excellence Program of the Ministry for Innovation and Technology.
I. L. acknowledges support from the European Research Council (ERC) under the European Union’s Horizon 2020 research and innovation programme grant agreement No. 771537. M. K. was supported by a ``Bolyai J\'anos'' grant of the HAS.

\newpage
\null
\newpage

\onecolumngrid

\centerline{\large\bf SUPPLEMENTARY MATERIAL}

\section{Bosonization approach}

	In this approach, 
	we consider an equilibrium 
	fermionic system with uniformly spaced one-particle energy levels. In the framework of  bosonization, the fermionic particle-hole excitations with respect to the ground state are represented as bosonic states. We assign thermal
	Boltzmann weights $e^{-\beta_\mathrm{eff}\,q\delta\varepsilon}$ to these states, where $\beta_\mathrm{eff}$ is an effective inverse temperature while $q\,\delta\varepsilon$ with $q=1,2,\dots$ measures the energy of the particle-hole excitation. Since these excitations are bosonic,
	for each $q$ we can have 
	$n_q=0,1,2,\dots$ arbitrarily many bosonic excitations with energy $qn_q\delta\varepsilon$. In the characteristic function each of them carries a contribution of 
	$e^{i\tilde u qn_q}$, so we have 
	\begin{equation}\label{eqGeff}
	\begin{split}
		G_{\tilde t}^\text{Bose}\argu u&=\mathcal{N}^{-1}\sum_{n_1,n_2,\dots}e^{-\argu{\beta_\mathrm{eff}\delta\varepsilon-i\tilde u}n_1}e^{-
		\argu{\beta_\mathrm{eff}\delta\varepsilon-i \tilde u}2n_2}e^{-\argu{\beta_\mathrm{eff}\delta\varepsilon-i\tilde u}3n_3}\dots=\mathcal{N}^{-1}
		\prod_{q=1}^\infty\sum_{n_q=0}^\infty e^{-qn_q\argu{\beta_\mathrm{eff}\delta\varepsilon-i\tilde u}}\\
		&=\mathcal{N}^{-1}\prod_{q>0}\frac{1}{1-e^{-q\argu{\beta_\mathrm{eff}\delta\varepsilon-i
		\tilde u}}},
		\end{split}
	\end{equation}
	where $\mathcal{N}=\prod_{q>0}\left[1-e^{-q\argu{\beta_\mathrm{eff}\delta\varepsilon}}\right]^{-1}$ so that $G_\mathrm{eff}(0,T_\mathrm{eff})=1$.
	Exponentiating Eq. \eqref{eqGeff} and taking the continuum limit $\sum_{q>0}\rightarrow\int_0^\infty\mathrm dx$ we get:
	\begin{equation}
		G_{\tilde t}^\text{Bose}\argu u\approx \mathcal{N}^{-1}e^{-\int_0^\infty\mathrm dx\ln\argz{1-e^{-\argu{\beta_\mathrm{eff}-iu}
		x}}}=e^{\frac{\pi^2/6}{\beta_\mathrm{eff}-i\tilde u}-\frac{\pi^2/6}{\beta_\mathrm{eff}}}\,.
	\end{equation}
	The Fourier transform can be computed analogously to the the mean field treatment above with the result
	\begin{equation}
		P_{\tilde t}^\text{Bose}\argu{w}\approx e^{-\frac{\pi^2}{6\beta_\mathrm{eff}}}\left[\frac{\pi}{\sqrt 6}e^{-\beta_\mathrm{eff}w}
		\frac{\mathrm{I}_1\argu{\pi\sqrt{\frac{2}{3}w}}}{\sqrt w}+\delta\left(w\right)\right]\,.
	\end{equation}

\section{Mean field approach}

In this section we provide some details about the mean field theory calculations and the resulting analytic expressions.

\subsection{Probability of adiabaticity}

Within the mean field approach, the probability of each many-body configuration takes the form of the product of independent Bernoulli weights of $M$ occupied and $N-M$ empty sites. In order to simplify calculations and without any loss of generality we consider the case of $M=N/2$:
\begin{equation}
\begin{split}
\label{eq:PMF}
P\left(\{n_k\}\right)&=\dfrac{1}{\mathcal{N}_t}\prod_{k=1}^N p_{k,t}(n_k)\;
\delta_{N/2=\sum_k n_k}\\
&=\dfrac{1}{\mathcal{N}_t}\int_{-\pi}^{\pi}\dfrac{{\rm d}\lambda}{2\pi}e^{i\lambda\sum_{k=1}^N\left(n_k -1/2 \right)}\prod_{k=1}^N p_{k,t}(n_k)\,,
\end{split}
\end{equation}
where the particle number conservation is taken into account by the Kronecker-delta for which we used a standard integral representation. The Bernoulli weights are
\begin{equation}
p_{k,t}(n_k) = n_k f_k(t)+(1-n_k)(1-f_k(t))\,,
\end{equation}
where $f_k\argu t=\argu{1-t_k\argu t}/2$ with $t_k(t)=\mathrm{erf}\left(\Delta k/\sqrt{4\widetilde D\tilde t}\right)$ and $\Delta k = k-M-1/2$ is measured from the Fermi-level.
Finally, the time-dependent normalization factor is the sum of all possible many-body probabilities:
\begin{equation}
\begin{split}
	\mathcal N_t\equiv\sum_{\left\{n_k\right\}}P(\left\{n_k\right\})&=\int_{-\pi}^\pi\frac{\mathrm d\lambda}{2\pi}\prod_{k=1}^N\left[e^{i
	\lambda/2}f_k(t)+e^{-i\lambda/2}(1-f_k(t))\right]=\int_{-\pi}^\pi\frac{\mathrm d\lambda}{2\pi}\prod_{k=1}^N\left[\cos(\lambda/2)-i
	\sin(\lambda/2)t_k(t)\right]\\
    &=\int_{-\pi}^\pi\frac{\mathrm d\lambda}{2\pi}\prod_{\Delta k>0}\left[\cos^2(\lambda/2)+\sin^2(\lambda/2)t_k(t)\right]\,.
    \end{split}
\end{equation}
Writing the above expression as the exponential of its logarithm, approximating the resulting sum by an integral and performing a saddle point approximation around $\lambda=0,$ we obtain for large enough values of $\langle w\rangle = \widetilde D\tilde t \gg1$:
\begin{equation}
\begin{split}
	\mathcal N_t&\approx\int_{-\pi}^\pi\frac{\mathrm d\lambda}{2\pi}\exp\left[\int_0^\infty\mathrm dx\log\left(\cos^2\lambda/2+t_x(t)\sin^2\lambda/
	2\right)\right]\\
	&\approx \int_{-\pi}^\pi\frac{\mathrm d\lambda}{2\pi}\exp\left[-\int_0^\infty\mathrm dx\lambda^2/4(1-t_x(t))\right]=\left(8\pi\avgw\right)^{-1/4}\,.
	\end{split}
\end{equation}
The probability of adiabaticity then reads 
\begin{align}
P_\text{ad}(\tilde t) &= \dfrac{1}{\mathcal{N}_t}\prod_{\Delta k<0}  f_{k}(t) \prod_{\Delta k>0}\big(1- f_{k}(t)\big)
\nonumber
 \\
&\approx 
\dfrac{1}{\mathcal{N}_t} e^{2\sqrt{4\avgw}\int_0^\infty\! {\rm d}x \log[(1+\mathrm{erf}(x))/2]}={(8\pi\widetilde D\tilde t)^{1/4}}\,{e^{-C\sqrt{\widetilde D\tilde t}}}={(8\pi\avgw)^{1/4}}\,{e^{-C\sqrt{\avgw}}} 
\nonumber
\end{align}
with $C\approx1.35$.

\subsection{Variance of work}

\bigskip

For $\langle w\rangle\gg 1$, we approximate the variance of the work by 
neglecting the fluctuations of the energy levels~\cite{vonDelft_KondoBox_paper}, 
 $\varepsilon_k(t) \to  \Delta k  \,\delta\varepsilon$,
but incorporating the fluctuations of the occupation numbers.
For a given realization of ${\cal H}(t)$, this leads to the estimate
\begin{equation}
    \delta w^2 (t) \approx 
    \Big\langle \big( \sum_{k = 1}^{N} \Delta k\,  \hat n_{k,t}\big)^2\Big\rangle 
    - \Big\langle \sum_{k = 1}^{N} \Delta k\,  \hat n_{k,t} \Big\rangle^2,
    \nonumber
\end{equation}
where $\langle\dots\rangle$ denotes quantum average. Separating the diagonal terms, 
the RM average $\langle\delta w^2(t)\rangle_{\rm RM}$ can be written as
\begin{equation}
    \langle\delta w^2(t)\rangle_{\rm RM}  \approx 
   \sum_{k }\Delta k^2\,  \big\langle \big\langle \delta  \hat n_{k,t}^2\big\rangle  \big\rangle_{\rm RM}
    +  \sum_{k \neq k'}  \Delta k\,\Delta k'
     \big\langle\big\langle \delta  \hat n_{k,t}\delta  \hat n_{k',t}
     \big\rangle \big\rangle_{\rm RM} \,,
\label{eq:variance}
\end{equation}
where $\delta  \hat n_{k,t} \equiv   \hat n_{k,t} -\langle   \hat n_{k,t}\rangle$ is the deviation 
of the occupation number from the mean value. 
As the  $\hat n_{k,t}$ behave as binary random variables, the averages in the first term  
are given by
$ \langle \langle\delta  \hat n_{k,t}^2\rangle\rangle_{\rm RM} = f_k(t) \big(1-  f_k(t) \big)$. 
The correlators in this equation can be expressed in terms of the amplitudes $\alpha_k^m(t)$
 as  $\langle \delta \hat n_{k,t}\delta \hat n_{k',t}\rangle =
-  \big|\sum_{m=1}^{N/2} {\alpha_k^m (t)}^* \alpha_{k'}^m(t)\big|^2$.
The negativity of this correction implies that the level occupations are
\emph{anticorrelated}, as follows from particle number conservation. 

Neglecting this correction for the moment and replacing sums by integrals, we arrive at the estimate
\begin{equation}
 \langle  \delta w^2 (t)\rangle_{\rm RM}  \approx 
\int_{-\infty}^{\infty} {\rm d}x\, x^2\, \frac{1 - {\rm erf}^2(x / \sqrt{4\widetilde{D}\tilde{t}})}{4}  \sim \tilde{t}^{3/2}\,,
\nonumber
\end{equation}
yielding $ \langle\delta w^2 (t)\rangle\sim\langle w\rangle^{3/2}$. We thus recovered the observed behavior, however, the prefactor turns out to be incorrect. A more careful mean field calculation shows that the occupation number correlations (related to fermion number conservation)  
 cannot be neglected but they also turn out to give a (smaller) $\sim \tilde t^{3/2}$ contribution, thus altering the prefactor but keeping the overall scaling the same.

\subsection{Distribution of work}
The characteristic function of the distribution of work  can be expressed as
    \begin{equation}
    \begin{split}
        G^\text{MF}_{\tilde t}\argu u&=\left\langle e^{iu\sum_k\Delta k\,\delta\varepsilon \,n_k}\right\rangle_\mathrm{MF}e^{-i uE_\mathrm{GS}}=\sum_{\left\{n_k\right
        \}}P_{\left\{n_k\right\}}e^{i\tilde u\sum_k\Delta k(n_k-1/2)}\\
        &=\frac{1}{\mathcal N_t}\int_{-\pi}^\pi\frac{\mathrm d\lambda}{2\pi}\prod_{\Delta k}
        \argz{e^{i\argu{\lambda+\tilde u \Delta k}/2}f_k\argu t+e^{-i\argu{\lambda+\tilde u \Delta k}/2}f_{-k}\argu t}\\
        &=\frac{1}{\mathcal N_t}\int_{-\pi}^\pi\frac{\mathrm d\lambda}{2\pi}\prod_{\Delta k<0}\argz{f_k(t)+e^{-i\argu{\lambda+\tilde u \Delta k}}f_{-k}\argu t}
        \prod_{\Delta k>0}\argz{f_{-k}(t)+e^{i\argu{\lambda+\tilde u \Delta k}}f_k\argu t}\\
        &\approx\frac{1}{\widetilde{\mathcal N}_t}\int_{-\pi}^\pi\frac{\mathrm d\lambda}{2\pi}
        \exp\argz{\int_0^\infty\mathrm dx\ln\argu{1+h^2_x\argu{u,t}+2h_x\argu{u,t}\cos\argu\lambda}},
    \end{split}
   \end{equation}
where we introduced the scaled variable  $\tilde u=u\,\delta\varepsilon$ and the notation $h_k(u,t)=\frac{f_k(t)}{{f_{-k}(t)}}e^{i\tilde u \Delta k}.$
   Here $\langle\dots\rangle_\mathrm{MF}$ denotes averaging over the mean field many-body probabilities and $\widetilde{\mathcal N}_t$ a modified 
    normalization constant. As numerics revealed, for large enough injected works $\avgw\gg1$ neglecting particle number 
    conservation does not introduce big errors provided we subtract the pure particle-hole excitations with respect to the ground state:
    \begin{equation}    
            G^\text{MF}_{\tilde t}\argu u\approx\frac{1}{\widetilde{\mathcal N}_t}\Bigg\{e^{2\int_0^\infty\mathrm dx\ln\argu{1+h_x\argu{t,u}}}
           -2\argz{e^{\int_0^\infty\mathrm dx\ln\argu{1+h_x\argu{t,u}}}-1}\Bigg\},
    \end{equation}
    where the first term is the $\lambda=0$ saddle-point solution of the integral expression, while the second part substracts the contributions coming from the pure particle-hole excitations.
Here the integrals can be approximated as
    \begin{equation}
        2\int_0^\infty\mathrm dx\ln\argz{1+h_x\argu{u,t}}\approx\frac{c^2_w}{\frac{c_w}{\sqrt{\avgw}}-iu}
    \end{equation}
yielding
    \begin{equation}        
        G^\text{MF}_{\tilde t}\argu u\approx\frac{1}{\widetilde{\mathcal N}_t}\args{e^{\frac{c^2_w}{\frac{c_w}{\sqrt{\avgw}}-
        iu}}-2\argz{e^{\frac{c^2_w/2}
        {\frac{c_w}{\sqrt{\avgw}}-iu}}-1}}
    \end{equation}
    with $c_w=\frac{3\sqrt{2\pi}}{5}$ chosen such that the characteristic function correctly reproduces the first two cumulants of work in the saddle 
    point solution. Now this expression can be Fourier transformed exactly as
   \begin{equation}
      \begin{split}
        \int_{-\infty}^\infty\frac{\mathrm du}{2\pi}e^{-iuw}e^{\frac{c^2_w}{\frac{c_w}{\sqrt{\avgw}}-iu}}&=\sum_{n=0}^\infty\frac{c^{2n}_w}{n!}\int_{-
        \infty}^\infty\frac{\mathrm du}{2\pi}\frac{e^{-iuw}}{\argu{\frac{c_w}{\sqrt{\avgw}}-iu}^n}=e^{-\frac{c_w}{\sqrt{\avgw}}w}\sum_{n=1}^\infty
        \frac{c^{2n}_ww^{n-1}}{n!\argu{n-1}!}+\delta\argu w\\
        &=e^{-\frac{c_w}{\sqrt{\avgw}}w}c_w\frac{I_1\argu{2c_w\sqrt w}}{\sqrt{w}}+\delta\argu w
      \end{split}
    \end{equation}
which leads to the approximate analytic expression
    \begin{equation}    
        P^\text{MF}_{\tilde t}\argu w\approx e^{-c_w\sqrt{\avgw}}\argz{e^{-\frac{c_w}{\sqrt{\avgw}}w}c_w\argu{\frac{I_1\argu{2c_w\sqrt w}}
        {\sqrt{w}}-\frac{I_1\argu{2c_w\sqrt{ w/2}}}{\sqrt{w/2}}}+\delta\argu w}\,.
    \end{equation}

\section{Energy space diffusion}

\begin{figure}[t]
\centering
\includegraphics[width=0.46\textwidth]{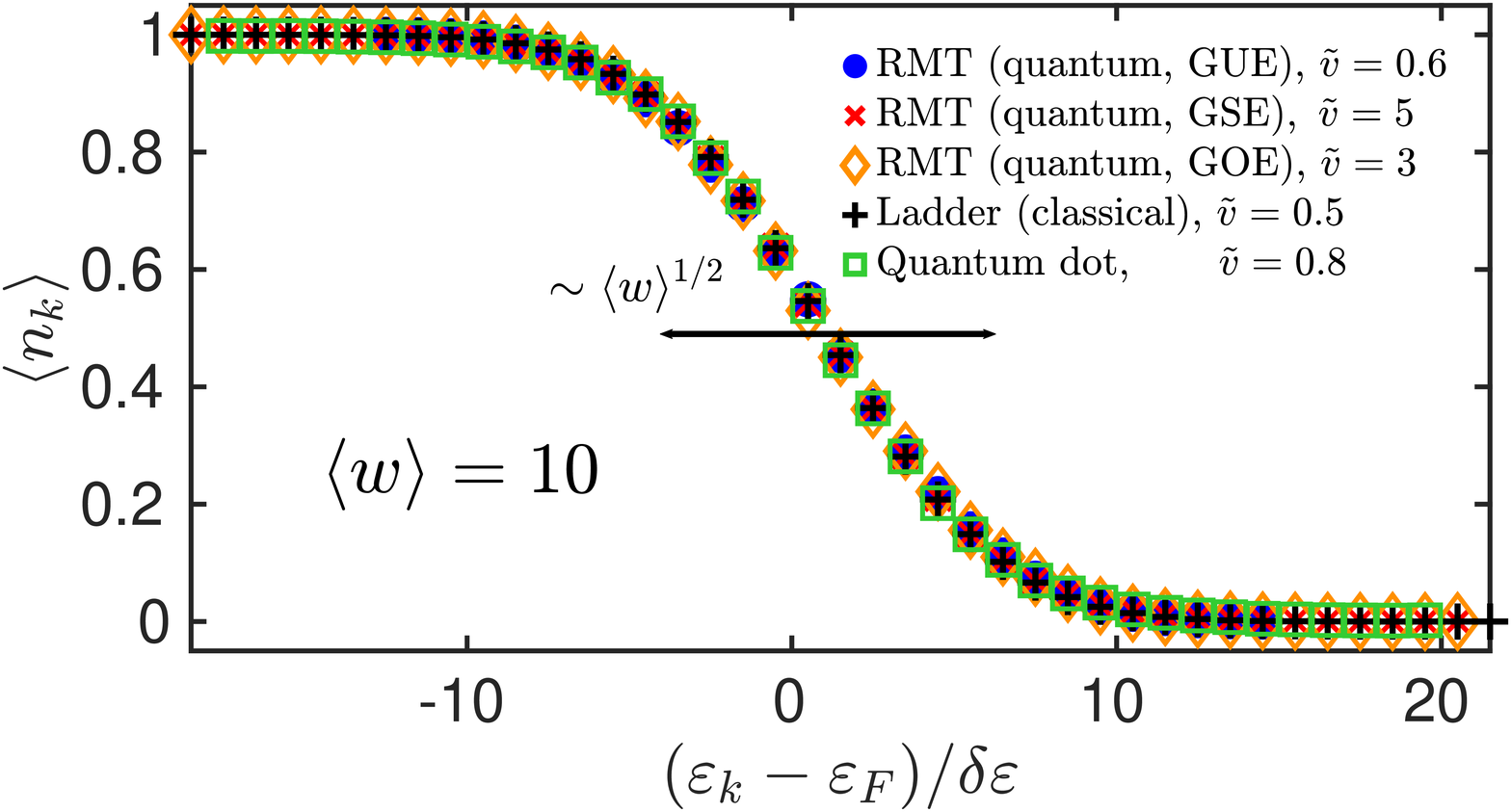}
\hfill
\includegraphics[width=0.48\textwidth]{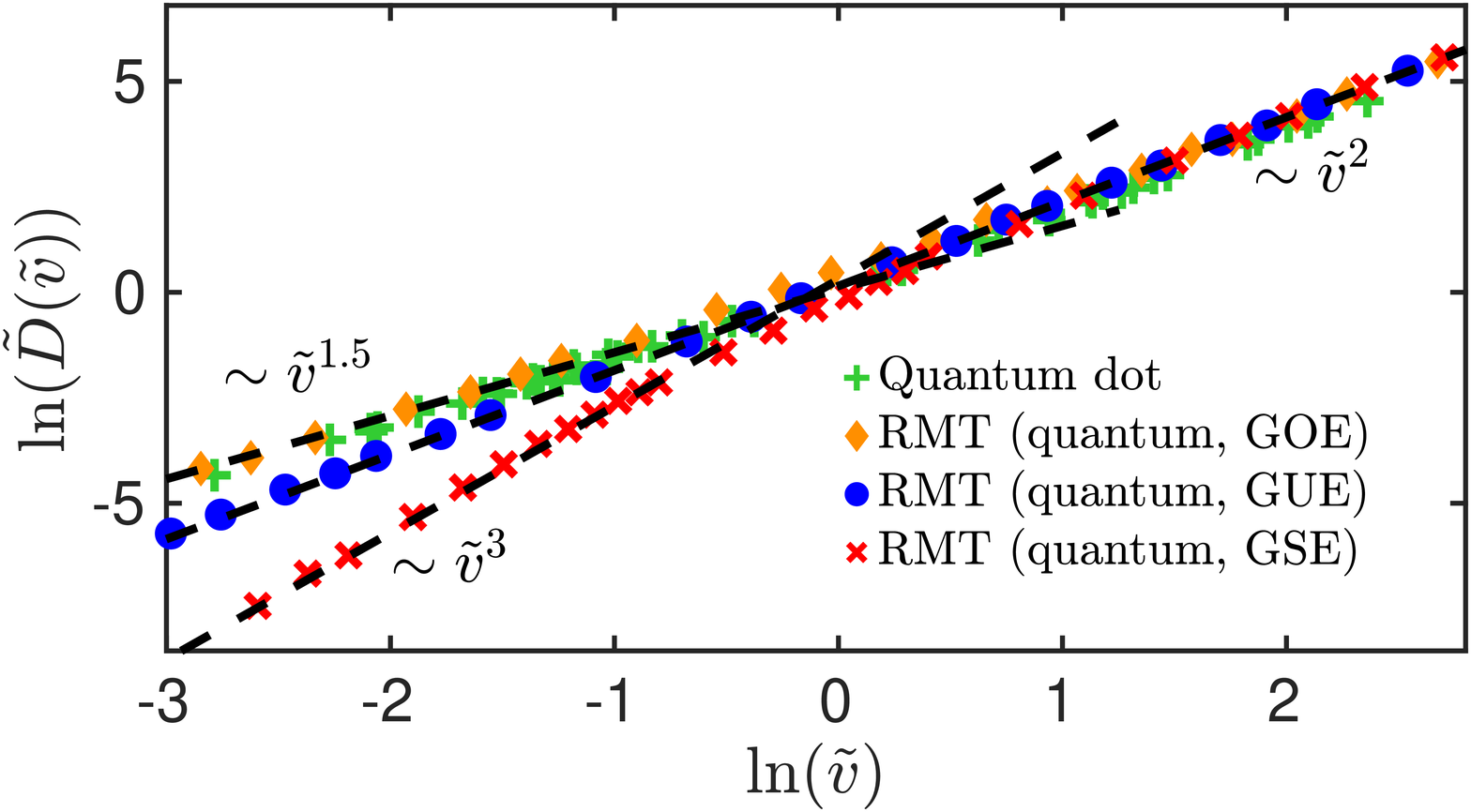}
\caption{Energy space diffusion.  {\it Left:} Average occupations of instantaneous single particle eigenstates for the three RMT ensembles (GOE, GUE, GSE) and the quantum dot model compared to the classically obtained results within the ladder model. All the five curves collapse onto a single universal, diffusively broadening profile given by $\argz{1-\mathrm{erf}\argu{\Delta k/\sqrt{4\avgw}}}/2$. 
{\it Right:} Velocity dependence of the diffusion constant. For slow quenches it has an anomalous power-law behavior, $\widetilde D(\tilde v\lesssim1)\sim\tilde v^{\beta/2+1}$, while for fast quenches it grows quadratically and with the same prefactor for the RMT ensembles. The quantum dot displays a similar behavior as the GOE ensemble in the two limiting cases, with a slightly different prefactor. \label{fig:diffus}}
\end{figure}

In this section we demonstrate that the energy level occupations exhibit a diffusive profile, meaning that particle-hole excitations happen dominantly in a window growing as $\sim\avgw^{1/2}$, for all the random matrix ensembles as well as for the ``ladder model'' and the disordered quantum dot. The left panel of Fig. \ref{fig:diffus} shows that for large enough average work the mean level occupation for of all three RMT ensembles (GOE, GUE, GSE) follows a single universal curve identical to those of the quantum dot model up to high precision and it is also perfectly described by the ladder model. Numerical calculations were made for $\sim 5\times 10^3$ disorder realizations both for RMT and the disordered quantum dot, for $N=40,28,40$ for the three ensembles, respectively and for parameters $L=38,\sigma=1.75J,\alpha=75J$ and with 427 particles in the case of the quantum dot.

The right panel of Fig. \ref{fig:diffus} shows the velocity dependence of the diffusion constant, $\widetilde D_\beta(\tilde v)$ for the three ensembles and the quantum dot model. We averaged over $\sim5\times10^3$ simulations, yielding smooth enough time-evolutions of average work to extract the diffusion constants. Parameters were chosen such that we avoid finite size effects and be in the diffusion regime. The rate of energy absorbed by the system exhibits an anomalous frequency dependence for slow quenches, $\widetilde D_\beta(\tilde v\lesssim1)\sim\tilde v^{\beta/2+1}$, while for fast processes becomes independent of the underlying symmetry class and grows quadratically, as it should in the case of a metal, $\widetilde D_\beta(\tilde v\gg1)\sim\tilde v^2$. The diffusion constant for the quantum dot shows the same power-law behavior as the GOE ensemble, albeit with a slightly smaller prefactor. 

\begin{figure}[t]
\centering
\includegraphics[width=0.7\textwidth]{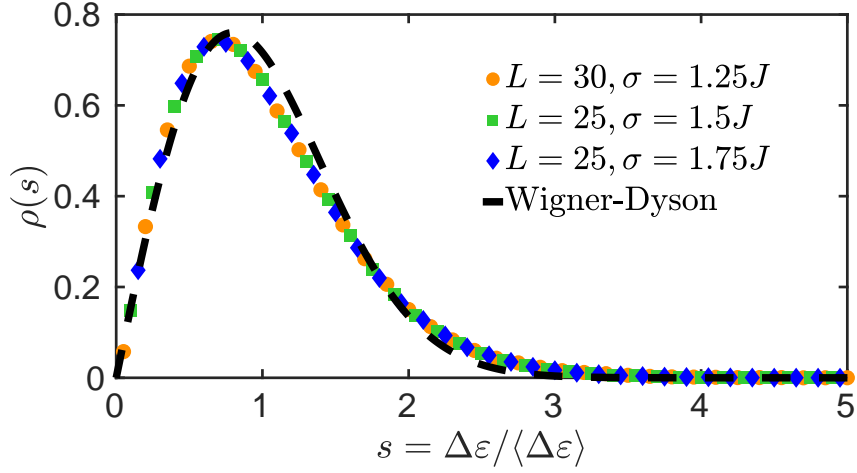}
\caption{Distribution of the distance between neighboring levels in the middle of the spectrum, $\Delta\varepsilon\equiv\varepsilon_{L^2/2+1}-\varepsilon_{L^2/2}$, normalized to unit mean, for the quantum dot at three different set of parameters, $L=30,\sigma=1.25J$, $L=25,\sigma=1.5J$, $L=25,\sigma=1.75J$ for the orange circles, green squares and blue diamonds, respectively. The potential strength is kept fixed, $\alpha=70J$ for all the three curves. The dashed line indicates the well-known Wigner--Dyson result, $\rho(s)\approx\frac{\pi}{2}s\,e^{-\frac{\pi}{4}s^2}$, obtained by Wigner's surmise describing the GOE case. For the numerical calculations we averaged over $\sim 5\times10^4$ disorder realizations which proved to yield smooth enough curves.\label{fig:WD}}
\end{figure}

Finally, we compare the level spacing distribution of the GOE ensemble and the disordered quantum dot. 
As shown in Fig. \ref{fig:WD}, 
the distribution of the distance of neighboring levels are well described by the analytical RMT result given by the Wigner surmise. Similar observations hold for the statistics of the the Landau--Zener parameters at the avoided level crossings in comparison with the RMT results of Ref.~[\onlinecite{wilkinson3}].

\end{document}